\documentclass[11pt,twoside]{article}  
\usepackage{adassconf}

\begin{document}   

%
%

\paperID{ }

%
%
%
%
\title{Massive Science with VO and Grids}


%

\author{Robert Nichol\altaffilmark{1}}
\affil{
Institute of Cosmology and Gravitation (ICG), 
Univ. of Portsmouth, Portsmouth, PO1 2EG, UK}

\author{Garry Smith\altaffilmark{2}}
\affil{Institute of Astronomy, School of Physics, University of Edinburgh, UK.} 

\author{Christopher Miller}
\affil{Cerro-Tololo Inter-American Observatory, NOAO, Casilla 603, LaSerena, Chile}

\author{Peter Freeman, Chris Genovese, Larry Wasserman,}
\affil{Dept. of Statistics, Carnegie Mellon University, Pittsburgh, PA-15213, USA}

\author{Brent Bryan, Alexander Gray\altaffilmark{3}, Jeff Schneider, Andrew Moore}
\affil{School of Computer Science, Carnegie Mellon University, Pittsburgh, PA-15213, USA}

\altaffiltext{1}{Email: bob.nichol@port.ac.uk}
\altaffiltext{2}{ICG, 
Univ. of Portsmouth, Portsmouth, PO1 2EG, UK}
\altaffiltext{3}{Georgia Tech College of Computing,
 801 Atlantic Drive,
 Atlanta, UK}


\contact{Bob Nichol}
\email{bob.nichol@port.ac.uk}

%
%
%
%
%

\paindex{Nichol, R. C.}

%
%

\authormark{Nichol, R. C., et al.}




\begin{abstract}          
  There is a growing need for massive computational resources for the
  analysis of new astronomical datasets. To tackle this problem, we
  present here our first steps towards marrying two new and emerging
  technologies; the Virtual Observatory (e.g, AstroGrid) and the
  computational grid (e.g. TeraGrid, COSMOS etc.). We discuss the
  construction of {\it VOTechBroker}, which is a modular software tool
  designed to abstract the tasks of submission and management of a
  large number of computational jobs to a distributed computer system.
  The broker will also interact with the AstroGrid workflow and
  MySpace environments. We discuss our planned usages of the {\it
    VOTechBroker} in computing a huge number of n--point correlation
  functions from the SDSS data and massive model-fitting of millions
  of CMBfast models to WMAP data. We also discuss other applications
  including the determination of the XMM Cluster Survey selection
  function and the construction of new WMAP maps.
\end{abstract}


\section{Introduction}

Over a petabyte of raw astronomical data is expected to be collected
in the next decade (see Szalay \& Gray 2001).  This explosion of data
also extends to the volume of parameters measured from these data
including their errors, quality flags, weights and mask information.
Furthermore, these massive datasets facilitate more complex analyses,
e.g.  nonparametric statistics, which are computationally intensive.
A key question therefore is: Can existing statistical software
scale-up to cope with such large datasets and massive calculations? We
address this question here by focusing on two exciting new
technologies, namely the Virtual Observatory (VO) and computational
grids.

\section{N--point Correlation Functions}

As a case study of the types of massive calculations planned for the
next generation of astronomical surveys and analyses, we discuss here
the galaxy n-point correlation functions. These have a long history in
cosmology and are used to statistically quantify the degree of spatial
clustering of a set of data points (e.g. galaxies). There are a
hierarchy of correlation functions, starting with the 2-point
correlation function, which measures the joint probability of a data
pair, as a function of their separation $r$, compared to a Poisson
distribution,

\begin{equation}
dP_{12} = N^2 dV_1\,dV_2 (1+\xi(r)), 
\end{equation}

\noindent where
$dP_{12}$ is the joint probability of an object being located in both
search volumes $dV_1$ \& $dV_2$, and $N$ is the space density of
objects. $\xi(r)$ is the 2-point correlation function and is zero for
a Poisson distribution. If $\xi(r)$ is positive, then the objects are
more clustered on scales of $r$ than expected, and vica versa for
negative values. The next in the series is the 3-point correlation
function, which is defined as,

\begin{equation}
dP_{123} = N^3 dV_1\,dV_2\,dV_3
(1+\xi_{12}(r_{12})+\xi_{23}(r_{23})+\xi_{13}(r_{13}) +
\xi_{123}(r_{12},r_{23},r_{13}) ),
\end{equation}

\noindent where $\xi_{12},\xi_{12},\xi_{12}$
are the 2-point functions for the three sides ($r_{12},r_{23},r_{13}$)
of the triangle and $\xi_{123}$ is the 3--point function. Likewise,
one can define a 4-point, 5-point etc., correlation function. The
reader is referred to Peebles (1980) for a full discussion of these
n-point correlation functions including their importance to cosmology
(see also the recent lecture notes of Szapudi 2005). We also refer the
reader to Landy \& Szalay (1993) and Szapudi \& Szalay (1998) for a
discussion of the practical details of computing the N--point
functions.

Naively, the computation of the n--point correlation functions scale
as $O(R^n)$, where $R$ is the number of data--points in the sample. As
one can see, even with existing galaxy surveys from the Sloan Digital
Sky Survey (SDSS), where $R\sim 10^6$--$10^7$, such correlation
functions quickly become untractable to compute. In recent years,
there has been a number of more efficient algorithms developed to beat
this naive scaling. For example, the International Computational
Astrostatistics (inCA; www.incagroup.org) group has developed a new
algorithm based on the use of the multi--resolutional KD-tree data
structure (mrKDtrees).  This software, known as {\it npt}, is publicly
available (www.autonlab.org), and has been discussed previously in
Gray et al. (2003), Nichol et al.  (2001) and Moore et al. (2000).
Briefly, mrKDtrees represent a condensed data structure in memory,
which is used to efficiently answer as much of any data query as
possible, i.e., pruning the tree in memory. The key advance of our
{\it npt} algorithm is the use of ``n'' trees in memory together to
compute an n--point function.

\begin{figure}[t]
\epsscale{.90}
\plotone{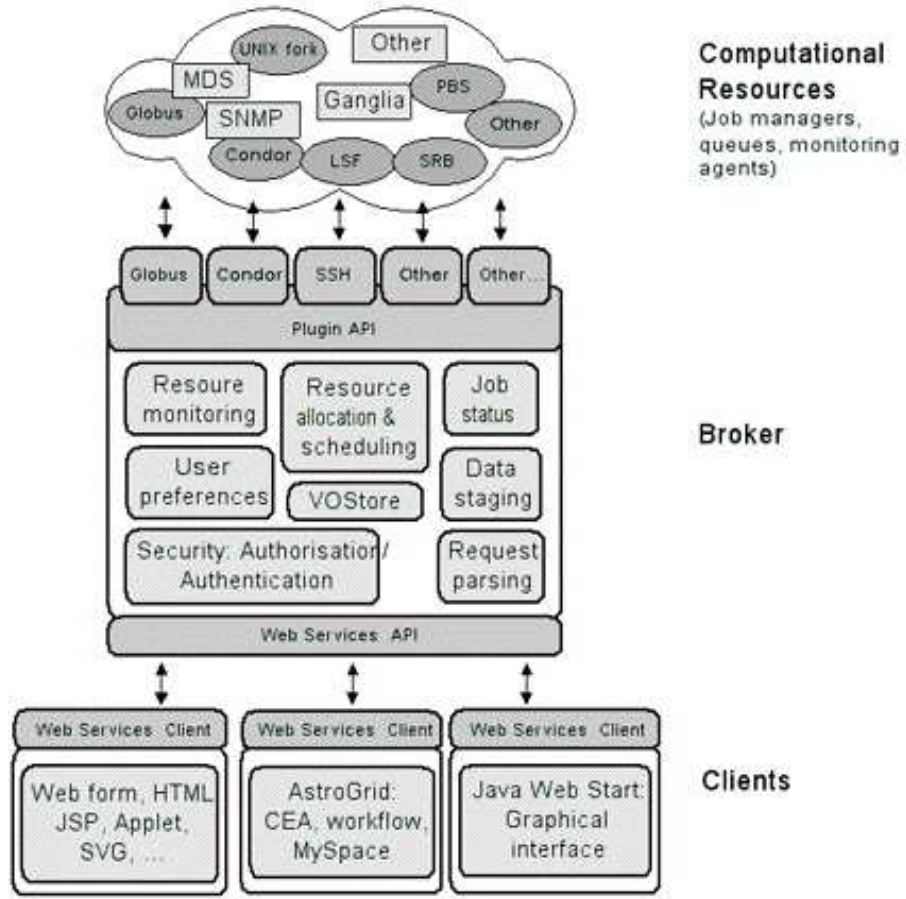}
\caption{The archtecture of the VOTechBroker\label{fig1}}
\end{figure}

\section{Computing Correlation Functions}

Even with an efficient algorithm, the computation of higher--order
correlation functions is intensive.  In detail, the n--point
correlation functions require a large number of sequential calls to
the {\it npt} code. These include computing the cross--correlation
between the real data (called $D$) and a random dataset (called $R$),
which is used to mimic the edge effects in the real data. As outlined
in Szapudi \& Szalay (1998), each estimation of a 3--point correlation
functions, for a given bin of triangle shapes (i.e.,
$r_{12}\pm\Delta_{r_{12}}$, $r_{23}\pm\Delta_{r_{23}}$,
$r_{13}\pm\Delta_{r_{13}}$, requires seven separate source counts over
the whole dataset, namely $DDD, DDR, DRR, RRR, DD, RR, DR$. Therefore,
if one wished to probe $\sim 10^2$ triangle configuration, then
$\sim10^3$ sequential {\it npt} jobs are required, each of which could
take several minutes to run.  This can rise rapidly if one wishes to
estimate errors on the n--point functions using either jack-knife
resampling (i.e., removing subregions of the data and then
re-computing the correlation functions), or a large ensemble of mock
catalogs (derived from simulations). Such computations are well-suited
to large clusters or grid of computers.

In recent years, we have used resources like TeraGrid
(www.teragrid.org) and COSMOS (www.damtp.cam.ac.uk/cosmos/) to perform
the computation of the n--point correlation functions for the SDSS
main galaxy sample and the SDSS LRG sample. Our experience shows that
the management and scheduling of such a large number of jobs on these
massive machines is laborious and tedious. To ease this problem, we
are working on {\it VOTechBroker}, which is a tool that joins two new
and emerging technologies; the VO and computational grids.

\section{VOTechBroker}

AstroGrid (www.astrogrid.org) is a PPARC-funded project to create a
working Virtual Observatory for UK and international astronomers.
AstroGrid works closely with other VO initiatives around the world
(via the International Virtual Observatory Alliance; IVOA) and is part
of the Euro--VO initiative in Europe. In particular, the work outlined
here has been performed as part of the EU--funded VOTech project,
which aims to complete the technical preparation work for the
construction of a European Virtual Observatory. Specifically, VOTech
is undertaking R\&D into data--mining and
visualization tools, which can be integrated into the emerging VO and
computational grid infrastructure. Therefore, VOTech will build upon
existing or emerging standards and infrastructure (e.g. IVOA standards
and AstroGrid middleware), as well as looking at standards from W3C
and GGF.

As part of the VOTech research, we are engaged in developing the {\it
  VOTechBroker}. The key design goals of the broker are to: {\it i)}
Remove the execution and management of a large number of jobs (like
{\it npt}) from the user in a transparent and reusable way; {\it ii)}
Accommodate different grid infrastructures (e.g. condor, globus etc.);
{\it iii)} Locate suitable resources on the grid and optimize the
submission of jobs; {\it iv)}
Monitor the status and success of jobs; {\it v)} Combine with AstroGrid
MySpace and workflow environments to allow easy management of job
submission and final results (as well as utilizing other algorithms
within the VO). In Figure \ref{fig1}, we show the schematic design of
the broker archtecture which illustrates the modular and ``plug-in''
design philosophy we have adopted. This is required as one of the key
requirements of {\it VOTechBroker} is that it should be straightforward to
add new algorithms, resources and middleware (e.g. a different job
submission tool or protocol).

We have implemented the core functionality of {\it VOTechBroker} and
are presently testing it by submitting $\sim10^4$ {\it npt} jobs on
both the UK National Grid Servise (www.ngs.ac.uk), COSMOS
supercomputer and a local condor pool of machines. The key ingredients
of the present {\it VOTechBroker} include GridSAM (an open-source job
submission and monitoring web servise from the London e-Science
Centre), the UK e-Science X.509 certificates, MyProxy (a repository
for X.509 Public Key Infrastructure security credentials), and the Job
Submission Description Language (JSDL; a standard description of job
execution requirements to a range of resource managers from the Global
Grid Forum). At present, the {\it VOTechBroker} provides a web-form
interface to just the {\it npt} algorithm discussed above but is
modular in design so other algorithms can be easily added via other
web--forms. Results from the {\it VOTechBroker} will soon be placed in a users
AstroGrid MySpace. In the near future, we will interface the broker
with other computational resources, e.g., TeraGrid (see below), and
the AstroGrid workflow.

\begin{figure}[t]
\plotone{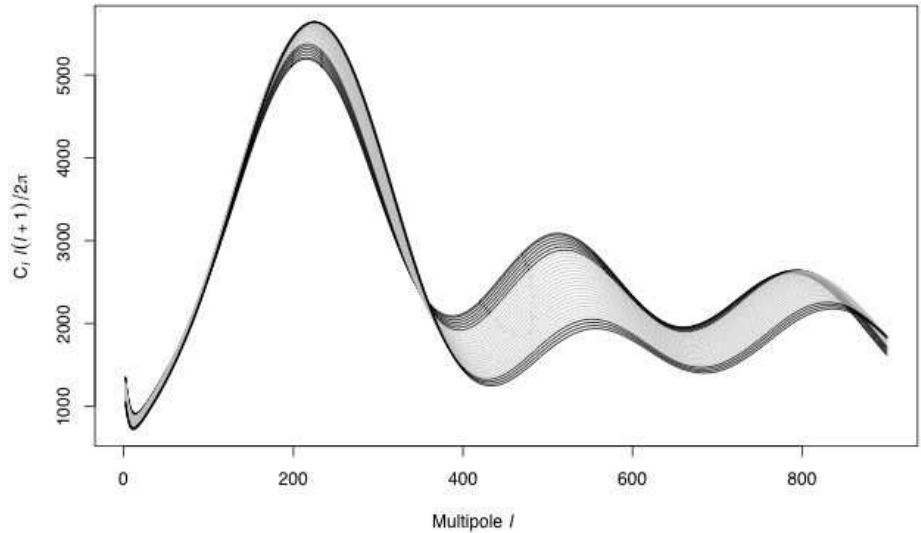}
\caption{Using CMBfast, we have varied $\Omega_b$ (baryon fraction) and determined which models lie within the 95\% confidence ball around $f(X_i)$. For this illustration, we have kept all other parameters in these CMBfast models fixed at their fiducial values. The gray models are within the confidence ball, while the others are outside the ball indicating they are ``bad fits'' to the data (at the 95\% confidence). We get an allowed range of $0.0169<\Omega_b<0.0287$.\label{fig2}}
\end{figure}

\section{Nonparametric Statistics}

In addition to the need for new statistical software that scales-up to
petabyte datasets, we also require new algorithms and computational
resources that exploit the emerging power of nonparametric statistics.
As discussed in Wasserman et al. (2001), such nonparametric methods
are statistical techniques that make as few assumptions as possible
about the process that generated the data.  Such methods are more
flexible than more traditional parametric methods that impose rigid
and often unrealistic assumptions. With large sample sizes,
nonparametric methods make it possible to find subtle effects which
might otherwise be obscured by the assumptions built into parametric
methods.

In Genovese et al. (2004), we discuss the application of nonparametric
techniques to the analysis of the power spectrum of anisotropies in
the Cosmic Microwave Background (CMB).  For example, one can ask the
simple question: How many peaks are detected in the WMAP CMB power
spectrum? This question is hard to answer using parametric models for
the CMB (e.g. CMBfast models) as these models possess
multiple peaks and troughs, which could potentially be fit to noise
rather than real peaks in the data.  To solve this, we have performed
a nonparametric analysis of the WMAP power spectrum (Miller et al.
2003), which involves explaining the observed data ($Y_i$) as
$Y_i=f(X_i) + c_i$ where $f(X_i)$ is a orthogonal function (expanded
as a cosine basis $\beta_i{\rm cos}(i\pi X_i)$) and $c_i$ is the
covariance matrix. The challenge is to ``shrink'' $f(X_i)$ to keep the
number of coefficients ($\beta_i$) to a minimum. We achieve this using
the method of Beran (2000), where the number of coefficients kept is
equal to the number of data points. This is optimal for all smooth
functions and provides valid confidence intervals. We also use
monotonic shrinkage of $\beta_i$, specifically the nested subset
selection (NSS). The main advantage of this methodology is that it
proves a ``confidence ball'' (in N dimensions) around $f(X_i)$,
allowing non-parametric interferences like: Is the second peak in the
WMAP power spectrum detected? In addition, we can test parametric
models against the ``confidence ball'' thus quickly assessing the
validity of such models in N dimensions. This is illustrated in Figure
\ref{fig2}.

\section{Massive Model Testing}

We are embarked on a major effort to jointly search the 7--dimensional
cosmological parameter--space of $\Omega_m,\Omega_{DE},\Omega_b,\tau$,
neutrino fraction, spectral index and H$_0$ using parametric models
created by CMBfast and thus determine which of these models fit within
the confidence ball around our $f(X_i)$ at the 95\% confidence limit.
Traditionally, this is done by marginalising over the other parameters
to gain confidence intervals on each parameter separately. This is a
problem in high-dimensions where the likelihood function can be
degenerate, ill-defined and under-identified. Unfortunately, the
nonparametric approach is computational intense as millions of models
need to searched, each of which takes $\simeq3$ minute to run.

\begin{figure}
\plotone{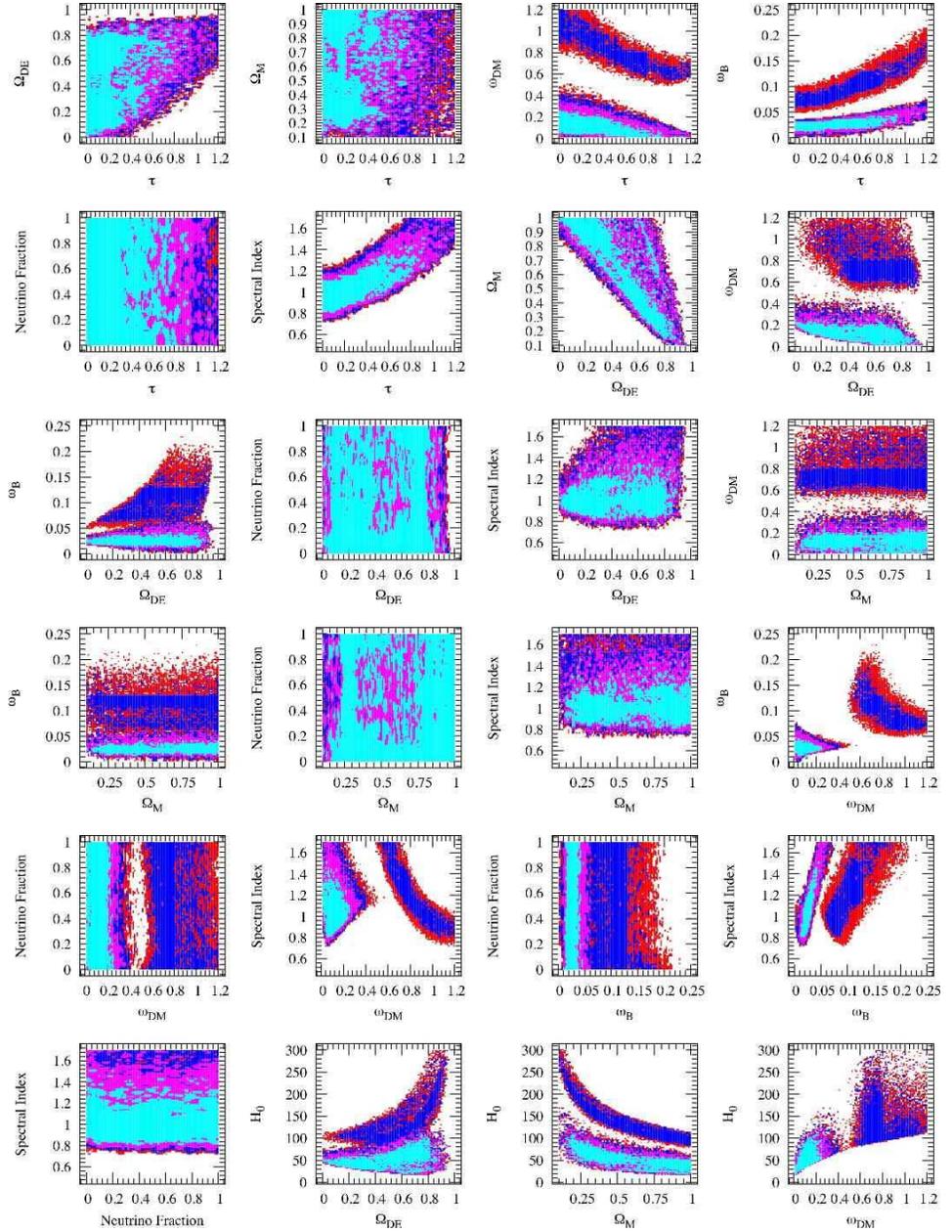}
\caption{The results of our 7--dimensional parameter search using 1.2 million models from CMBfast. The light blue (or lightest shading for greyscales) color are models excluded at the 34\% level. The purple (or mid-grade shading) are models excluded by the 68\% confidence ball and the red is the 95\% confidence ball \label{fig3}}
\end{figure}

To mitgate this problem, we have developed an intelligent method for
searching for the surface of the confidence ball in high-dimensions
based on Kriging. Briefly, kriging is a method of interpolation which
predicts unknown values from data observed at known locations (also
known as Gaussian process regression, which is a form of Bayesian
inference in Statistics). There are many different metrics for
evaluating the kriging success including variance and entropy, yet we employ the ``Straddle'' method which picks new test
points based both on the overall distance from other searched points
and are predicted to be near the boundary. We have also developed a
heuristic algorithm for searching for ``missed peaks'' in the
likelihood space by searching models along the path jointing already
detected peaks. We find no ``missed peaks'', which
illustrates our kriging algorithm is effective in finding the surface
of the confidence ball in this high dimensional space.

We have distributed the CMBfast model computations over a local condor
pool of computers. In Figure \ref{fig3}, we show preliminary results
from this high-dimension search for the surface of the confidence ball
and present {\bf joint} 2D confidence limits on pairs of the
aforementioned cosmological parameters. These calculations represent
6.8 years of CPU time to calculate over one million CMBfast models. In
the near future, we will move this analysis to TeraGrid, using {\it
  VOTechBroker}, and plan 10 million models to fully map the surface
of the confidence ball. We will also make available a Java--based web
servise for accessing these models, and the WMAP confidence ball, thus
allowing other users to rapidly combine their data with our WMAP
constraints e.g., doing a joint constraint from LSS and CMB data. We
are also working on possible convergence tests, and visualization tools
within VOTech, to access this high-dimensional data.

\section{Other Applications}

We present here two other examples of where massive computations are
needed and could greatly benefit from the {\it VOTechBroker}. First,
the XMM Cluster Survey (XCS) is a project dedicated to uniformily
analysing all XMM pointings in search of clusters of galaxies i.e.,
extended X-ray sources. Dedicated software has been written to find
clusters and a cluster target list created. One of the key components
for the cosmological analysis of the XCS is the selection function,
which is the efficiency (and therefore, the effective volume of the
survey) in finding clusters as function of cluster properties (cluster
profile, redshift, temperature) and observational constraints
(exposure time, background etc.). The most comprehensive method of
determining such selection functions is via extemsive Monte Carlo
simulations i.e., adding fake clusters to the real data and measure
the efficiency in re-detecting them (see Adami et al.  2000 for such
simulations for the SHARC survey). We estimate that over a million
simulations will be required to confidently estimate the XCS selection
function because of the large number of parameters involved. This
translates to over 4 years of CPU time to complete, but could be
trivially parallelized over a large grid of computers using {\it
  VOTechBroker}.

Next, we discuss the our independent analysis of the WMAP time-stream
data to create CMB maps of the sky (see Freeman et al. 2005). Such
independent confirmation is important and allows use to apply the same
non-parametric techniques discussed above to the WMAP map-making
procedures, and thus assess their effect on the maps and power
spectrum analyses. We found small differences between our maps and
WMAP (10 microKelvin), which are primary because of the residual
dipole uncertainty and second-order terms in the Doppler shift (See
Freeman et al. 2005 for details). Each map would take a CPU day on a
single processor, but we have parallalized the map-making code to
exploit TeraGrid and can make maps in minutes.

\section{Summary}

We provide here several examples of massive astronomical data analyses
that require significant computational resources. Our plan is to
develop the {\it VOTechBroker} to provide a power framework within
which such analyses can be performed. As discussed, the main goals of
the {\it VOTechBroker} are to abstract from the user (either a person
or another program) the complexities of job submission and management
on computational grids, as well as being a modular ``plug--in'' design
so other algorithms and software can be easily added. Finally, we plan
to integrate {\it VOTechBroker} into the AstroGrid workflow and
MySpace environments, so it becomes a natural repository for a host of
advanced statistical algorithms than scale-up in preparation for
petabyte-scale datasets and analyses.

\acknowledgments

We thank all our collaborators and colleagues in inCA, VOTech,
AstroGrid, SDSS and VO projects. The work presented here was partly
funded by NSF ITR Grant 0121671 and through the EU VOTech and Marie
Curie programs. RCN thanks the organisers of the ADASS meeting for
their invitation. GS thanks the VOTech and University of Edinburgh for
his funding (see eurovotech.org for details).

\end{document}